\documentclass[aps,twocolumn,superscriptaddress,showpacs,floatfix]{revtex4}
\usepackage{amsmath}
\usepackage{amsbsy}
\usepackage{amssymb}
\usepackage[dvipdfmx]{graphicx}
\usepackage{epsfig}
\usepackage{color}
\usepackage[normalem]{ulem}
\usepackage{subfigure}
\usepackage{verbatim}
\usepackage{psfrag}

\def\bse{\begin{subequations}}
\def\ese{\end{subequations}}

\newcommand{\fI}{f_{\mu}}
\newcommand{\fA}{f_{\mu}^{\rm A}}
\newcommand{\HI}{H_{\mu}}
\newcommand{\HA}{H_{\mu}^{\rm A}}
\newcommand{\ave}[1]{\left\langle #1 \right\rangle}

\begin{document}

\title{Trapping scaling for bifurcations in Vlasov systems}%

\author{J. Barr\'e}
\affiliation{Laboratoire J.-A. Dieudonn\'e, Universit\'e de Nice Sophia-Antipolis, CNRS, 06109 Nice, France,\\
 Institut Universitaire de France}
\author{D. M\'etivier}
\affiliation{Laboratoire J.-A. Dieudonn\'e, Universit\'e de Nice Sophia-Antipolis, CNRS, 06109 Nice, France.}
\author {Y.Y. Yamaguchi}
\affiliation{Department of Applied Mathematics and Physics, Graduate School of Informatics, Kyoto University, Kyoto 606-8501, Japan}

\date{\today{}}

\begin{abstract}
We study non oscillating bifurcations of non homogeneous steady states of the Vlasov equation, a situation occurring in galactic models, or for Bernstein-Greene-Kruskal modes in plasma physics. Through an unstable manifold expansion, we show that in one spatial dimension the dynamics is very sensitive to the initial perturbation: the instability may saturate at small amplitude -generalizing the "trapping scaling" of plasma physics- or may grow to produce a large scale modification of the system. Furthermore, resonances are strongly suppressed, leading to different phenomena with respect to the homogeneous case. These analytical findings are illustrated and extended by direct numerical simulations with a cosine interaction potential.\\
\end{abstract}
%

\maketitle

\section{Introduction}
The Vlasov equation models the dynamics of a large number of interacting particles,
when the force acting on them is dominated by the mean-field.
It is a fundamental equation of plasma physics and galactic dynamics,
but also describes a number of other 
physical systems, such as wave-particles interactions
\cite{Elskens_book},
free electron lasers~\cite{Bonifacio},
certain regimes of nonlinear optics~\cite{Picozzi_review},
and wave propagation in bubbly fluids~\cite{Smereka}.
Understanding its qualitative behavior has proved to be
a formidable challenge, from the first linear computations of 
Landau~\cite{Landau46} to the most recent mathematical
breakthroughs~\cite{MouhotVillani}.
We address here a problem of bifurcations in these systems.

The study of bifurcation of \emph{homogeneous} stationary solutions
to the Vlasov equation (when particles are homogeneously distributed in space,
with a certain velocity distribution) has a long and interesting history,
mainly related to plasma physics;
a detailed account can be found in~\cite{Balmforth_review}.
One of the main question is to determine the asymptotic behavior of the system,
when started close to an unstable stationary state.
The basic mechanism, ultimately responsible for the instability saturation,
is the \emph{resonance} phenomenon: particles with a velocity close to
that of the growing mode are strongly affected by the small perturbation.
This was qualitatively recognized in the 1960's~\cite{Baldwin64,ONeil65},
leading these authors to predict the famous "trapping scaling",
that is the asymptotic electric field, after nonlinear saturation,
would be of order $\lambda^2$, with $\lambda$ the instability rate.
However, formal computations relying on standard nonlinear expansions
often led to the very different $\lambda^{1/2}$ scaling
(see discussion and refs in \cite{Balmforth_review}).
In the 1990's, Crawford's careful unstable manifold computations
finally recovered unambiguously the trapping scaling~\cite{Crawford}.
Yet, his expansion is plagued by strong divergences
in the $\lambda \to 0^{+}$ limit.
Soon after, Del-Castillo-Negrete derived an infinite dimensional
reduced model for the instability development,
using different techniques~\cite{DelCastilloNegrete}.
This is currently the best understanding we have,
and shows a striking degree of universality.

Understanding the non homogeneous case is further challenging,
and has a practical importance for various physical systems.
A first key example comes from astrophysics.
Radial Orbit Instability destabilizes a spherical self-gravitating system
when the number of low angular momentum particles, usually stars, grows. 
The fate of the initially unstable system
has been qualitatively and numerically studied in \cite{Palmer90}, providing a rare example of such a bifurcation study.
In the plasma physics context, Bernstein-Greene-Kruskal exhibited a large family of
exact solutions of the Vlasov-Poisson equation \cite{BGK},
now commonly called BGK modes. 
Some of these solutions are stable, others unstable \cite{Schwarzmeier79,Manfredi00,Lin05,Pankavitch14}.
The non linear behavior in case of instability is an open question.

To tackle the bifurcation in the non homogeneous case
we will consider in this article as a first step the case of a real unstable eigenvalue,
for a one dimensional system.
We will show that the number of particles around zero frequency is negligible
in the non homogeneous case
compared to the homogeneous case.
This is a crucial remark, since it implies that for non oscillating instabilities, the resonance phenomenon will be absent, or strongly suppressed.
Nevertheless, similarly to the homogeneous case, 
the unstable eigenvalue still bifurcates from a marginally stable continuous spectrum. 
Thus, one should expect a new type of bifurcation.
Our strategy is to attack this problem by revisiting Crawford's expansion,
in order to describe the dynamics on the unstable manifold.
The basic questions are: Does the instability saturate at small amplitude?
What is the analog of the "trapping scaling"?
Can we expect the same kind of universality as in the homogeneous context?
Our main results include:
i) A striking asymmetry on the unstable manifold: one branch leads to a small amplitude saturation, the 
other escapes far away from the original stationary point.  
ii) The $\lambda^2$ "trappping scaling", well known in the homogeneous case, generalizes 
for the small amplitude saturation; however, the reason for this scaling is very different. This 
result can also be reached by a different theoretical approach, following \cite{OgawaYamaguchi15}.
iii) Since these features depend on some generic physical characteristic of the problem, we expect 
our results to be generic, at least for 1D systems. 

The outline of the article is as follows:
Section \ref{sec:gen} deals with a general one-dimensional Vlasov equation.
We first explain formally how to obtain the unstable manifold expansion
and emphasize the main physical points (Section \ref{sec:gen1}), then present a more detailed computation (Section \ref{sec:gen2}).
In Section \ref{sec:cosine}, we restrict to a particular cosine interaction potential.
We are then able to perform more explicit computations,
and provide precise predictions, as well as push the computations at higher orders, which is crucial to understand the divergence pattern of the 
expansion (Section \ref{sec:higher_orders}).
We also show how some of these predictions can also be obtained
via the "rearrangement formula" idea,
recently introduced~\cite{OgawaYamaguchi15}; this provides an independent theoretical approach (Section \ref{sec:rearrangement}).
In Section \ref{sec:simulations}, we compare all previous predictions with direct numerical simulations. The choice of the cosine potential allows 
us to perform very precise computations.
Section \ref{sec:conclusion} is devoted to some final comments and open questions.

\section{Unstable manifold reduction (spatially periodic 1D systems)}
\label{sec:gen}
\subsection{General picture}
\label{sec:gen1}
The Vlasov equation associated with the one-particle
Hamiltonian $H[f]$ is 
\begin{equation}
    \partial_{t}f = \{f, H[f]\},
    \quad
    H[f](q,p,t) = p^{2}/2 + V[f](q,t)
\end{equation}
where the one-particle potential $V[f]$ is
defined by using the two-body interaction $v(q)$ as
\begin{equation}
    V[f](q,t) = \int v(q-q') f(q',p',t) dq'dp'
\end{equation}
and the Poisson bracket $\{\cdot,\cdot\}$ is 
\begin{equation}
    \{ u, v\} = \partial_{p}u \partial_{q}v - \partial_{q}u \partial_{p}v. 
\end{equation}
Denote by $f_{\mu}$ a family of unstable stationary states 
depending on the parameter $\mu$, and by $g$ a perturbation around $f_{\mu}$.
The equation for $g$ consists of 
a linear and a nonlinear parts:
\begin{equation}
    \partial_{t}g = L_{\mu}g + N[g],
\end{equation}
where
\begin{equation}
    L_{\mu}g = \{ g, H[f_{\mu}] \} + \{ f_{\mu}, V[g] \}~,~ N[g] = \{ g, V[g] \}.
\end{equation}
Let $L_{\mu}$ have an unstable \emph{real} nondegenerate
eigenvalue $\lambda$
depending on $\mu$, and $\Psi_{\lambda}$ be the corresponding eigenfunction.
We assume that the family $\{f_{\mu}\}$ has a critical point $\mu_{\rm c}$,
where the eigenvalue is $\lambda=0$,
and we will study the system in the limit $\mu\to\mu_{\rm c}$
which implies $\lambda\to 0^{+}$. 

Notice that the Hamiltonian structure implies that $L_{\mu}$ also has the eigenvalue $-\lambda$. In addition, by  
space translation symmetry when $f_\mu$ is homogeneous in space, $\lambda$ is in this case twice 
degenerate. When $f_\mu$ is non homogeneous, the symmetry is broken, 
and a neutral Goldstone mode $\Psi_N$ appears.

We use the standard $L^2$ Hermitian product, and introduce $L^{\dagger}_{\mu}$,
the adjoint operator of $L_{\mu}$, and $\widetilde{\Psi}_{\lambda}$ the
adjoint eigenfunction with eigenvalue $\lambda^{\ast}=\lambda$,
where $\lambda^{\ast}$ is the complex conjugate of $\lambda$.
We normalize it as  $\ave{ \widetilde{\Psi}_{\lambda}, \Psi_{\lambda} } = 1$,
where the scalar product is defined by
  \begin{equation}
    \ave{ f, g } = \int\! f^{\ast} g~dqdp,
  \end{equation}
and note $\Pi_{\mu}(\cdot)=\ave{\widetilde{\Psi}_\lambda,\cdot}\Psi_\lambda$
the projection onto the unstable eigenspace {\rm Span}\{$\Psi_\lambda$\}.
Depending on the behavior of $\Psi_{\pm \lambda}$ and $\Psi_N$ 
when $\lambda \to 0^{+}$, $\Pi_{\mu}$ can be singular in this limit. As will become 
clear later on, this is the origin of an important difference between homogeneous 
and non homogeneous cases. 

We are interested in the dynamics on the unstable manifold of $f_{\mu}$,
and expand $g$ as
\begin{equation}
    g(q,p,t) = A(t) \Psi_{\lambda}(q,p) + S(q,p,A(t)).
\end{equation}
The term $S$ parameterizes the unstable manifold,
and we have the estimation $S=O(A^{2})$ for small $A$
by assuming that the unstable manifold is tangent to the unstable eigenspace.
Note that this parametrization may be only local.

Performing the projection,
we obtain for $A(t)$ the following dynamical equation, expanded in powers of $A$:
\begin{equation}
    \label{eq:reduced-eq}
    \dot{A} = \lambda A + \ave{\widetilde{\Psi}_{\lambda}, N[f]} = \lambda A +a_{2}(\lambda) A^{2} + a_{3}(\lambda) A^{3} + \cdots,
\end{equation}
where the coefficient $a_{2}$ is
\begin{equation}
\label{eq:a2}
    a_{2}(\lambda) = \ave{\widetilde{\Psi}_{\lambda} , \{ \Psi_{\lambda}, V[\Psi_{\lambda}] \} }.
\end{equation}
In the homogeneous case, the unstable eigenspace is actually two-dimensional,
and the reduced equation involves $A$, which is complex, and $A^\ast$;
since an equation similar to \eqref{eq:reduced-eq} can be written for $|A|$,
we omit this slight difference here.
In this case we have by symmetry $a_{2l}=0$ for any $l\in\mathbb{N}^*$ and $a_{2l+1}\propto \lambda^{1-4l}$
\cite{Crawford}, this strong divergence preventing the truncation 
of the series~\eqref{eq:reduced-eq}. By contrast, in the non homogeneous case, $a_2\neq 0$; we show 
however below that a similar divergence pattern when $\lambda \to 0^+$ is expected.
Nevertheless, keeping term up to $A^2$, \eqref{eq:reduced-eq} looks like the normal form of a transcritical bifurcation; it is not a standard one 
however, because $\lambda$ is always positive, and $a_2$ diverges when $\lambda \to 0^+$. 
From the truncated equation, we can conjecture the general phenomenology:
\begin{itemize}
\item on the unstable manifold, in one direction the dynamics
is bounded and attracted by a stationary state close to $f_\mu$,
characterized by $A _\infty= -\lambda/a_2(\lambda)$
(notice that this state is stationary and stable for the truncated unstable manifold dynamics;
it may not be so for the unconstrained dynamics);
\item in the other direction, the dynamics leaves the perturbative regime,
and thus the range of validity of our analysis. 
\end{itemize}
Estimating $a_{2}(\lambda)$ in the $\lambda \to 0^+$ limit gives access to $A _\infty$; this requires more technical work performed in the next subsection. 
Note that the formal computations leading to \eqref{eq:reduced-eq} and \eqref{eq:a2} are essentially independent of the dimension; the following estimate for
$a_2(\lambda)$ is valid in 1D.

\subsection{Technical computation}
\label{sec:gen2}
We present now the explicit form of $a_{2}(\lambda)$, the order $A^2$ coefficient in the reduced dynamics,
for the non homogeneous case of a generic $1$-dimensional system. The computation is somewhat technical, but 
the physical message is rather simple, and summarized here:
\begin{itemize}
\item $a_2 \propto 1/\lambda$ when $\lambda \to 0^+$ 
\item the normalization condition requires
  $\widetilde{\Psi}_{\lambda}=O(1/\lambda)$,
  and this divergence of the amplitude in the neighborhood of $\lambda=0$
  is responsible for the divergence of $a_{2}$.
\item the integrals appearing in the scalar product
in \eqref{eq:a2} do not diverge when $\lambda \to 0^+$.
\end{itemize} 
The last two points are major differences with the homogeneous case
in which the amplitude of $\tilde{\Psi}_{\lambda}$ does not diverge,
and the resonances translate into pinching singularities
for the integrals analogous to \eqref{eq:a2} \cite{Crawford}. These pinching singularities are responsible for the 
divergences in the unstable manifold expansion.
 From these results, we can now conclude that the stationary state close to $f_\mu$
on the unstable manifold is characterized by $A_\infty \propto \lambda^2$,
which is a central result of this article.

\subsubsection{The spectrum matrix}
Let us turn to the explicit computation.
We first need to introduce quantities regarding the linearized problem; we follow a classical method \cite{Kalnajs77,Lewis79} and use here the notations of \cite{BOY11}. For preparation, we introduce the biorthogonal functions 
$\{d_{i}\}_{i\in I}$ and $\{u_{k}\}_{k\in K}$, on which we expand the density and potential respectively:
\begin{equation}
    \int f(q,p,t) dp = \sum_{i\in I} a_{i}(t) d_{i}(q),
\end{equation}
\begin{equation}
    V[f](q,t) = \sum_{k\in K} a_{k}(t) u_{k}(q).
\end{equation}
The functions $\{d_{i}\}$ and $\{u_{k}\}$ satisfy
\begin{equation}
    \int v(q-q') d_{k}(q') dq' = u_{k}(q),
\end{equation}
and
\begin{equation}
    (d_{i},u_{k}) = \int d_{i}(q)^{\ast} u_{k}(q) dq
    = \nu_{k}\delta_{ik},
\end{equation}
Since all the functions expanded are real, we can choose the $d_i$'s and $u_k$'s to be real functions; the $\nu_k$'s are then also real.

In a generic system the spectrum function
is a matrix denoted by $\Lambda_{\mu}(\lambda)$ \cite{Kalnajs77,Lewis79},
whose $(l,k)$ element is
\begin{equation}
    (\Lambda_{\mu})_{lk}(\lambda)
    = \nu_{k}\delta_{lk} - 2\pi \sum_{m\in\mathbb{Z}}
    \int \dfrac{im\nabla_{J}F_{\mu}}{\lambda+im\Omega}
    c_{k,m}c_{l,m}^{\ast} dJ,
\label{eq:Lambda}
\end{equation}
where we have introduced the action-angle variables $(J,\theta)$
for an integrable Hamiltonian system $H[f_{\mu}]$ with $1$ degree of freedom,
and assumed that $f_{\mu}$ depends only on $J$;
this is denoted by $f_{\mu}(q,p)=F_{\mu}(J)$.
The function $c_{k,m}(J)$ is defined by
\begin{equation}
    c_{k,m}(J) = \dfrac{1}{2\pi} \int u_{k}(q) e^{-im\theta} d\theta,
\end{equation}
and the frequency $\Omega(J)$ by $\Omega=dH[f_{\mu}]/dJ$.
We remark that $\det\Lambda_{\mu}(\lambda)=0$ implies $\lambda$
to be the eigenvalue of the linear problem.
The matrix's size is $\# K$,
the number of biorthogonal functions in the expansion of the potential; it is usually infinite, 
but only $2\times 2$ for the cosine potential introduced later.

\subsubsection{Properties of the spectrum matrix}
To analyze $\Lambda$, we need to specify
a little bit more the frequency $\Omega(J)$.
We assume that $\Omega(J)$ does not vanish at finite $J$, or does so only logarithmically. This includes the cases where the 
stationary potential $V[f_{\mu}](q)$ has a single minimum and is infinite for $|q|$ infinite (such as for 1D gravity), and the generic situation with periodic boundary conditions; indeed, in the latter situation, local minima of the stationary potential give rise to separatrices, on which the action is constant. At these specific values of the 
action $\Omega$ vanishes, but generically it does so only logarithmically (an example of this situation is given in the next section).
In particular, under the above assumption for $\Omega$, the functions $(\Lambda_{\mu})_{lk}(\lambda)$ defined in \eqref{eq:Lambda} are well defined for $\lambda=0$ (the integral over $J$ converges), and continuous for $\lambda \in \mathbb{R}$. Furthermore, the same reasoning applies to the derivatives of $(\Lambda_{\mu})_{lk}(\lambda)$: these functions are continuous for $\lambda \in \mathbb{R}$, even for $\lambda=0$. The fact that integrals over $J$ converge is the technical counterpart of a physical phenomenon: the absence, or weakness, of resonances for $\lambda=0$.

Furthermore, starting from \eqref{eq:Lambda}, changing the summation variable $m$ to $-m$, and using $c_{k,-m}(J)=c_{k,m}^{\ast}(J)$, it is not difficult to see that 
$\Lambda_\mu$ is a real matrix, and that $(\Lambda_{\mu})_{lk}(\lambda)=(\Lambda_{\mu})_{kl}(-\lambda)$. Hence $\Lambda_{\mu}(0^{+})$ is symmetric. This will be useful below.

\subsubsection{Comparison with the homogeneous case}
The properties of the spectrum matrix detailed in the above paragraph are in sharp contrast with the better known situation for a homogeneous stationary state. It is useful to highlight the comparison. We assume here periodic boundary conditions. For a homogeneous stationary state, a Fourier transform with respect to the space variable diagonalizes the spectrum matrix. The diagonal elements are \cite{Crawford}:
\begin{equation}
\label{eq:Lambda_hom}
\Lambda^{\rm{(hom)}}_k(\lambda) = 1-2i\pi k \hat{v}_k\int \frac{F'_0(p)}{\lambda+ikp}dp,
\end{equation}
where $F_0(p)$ is the momentum distribution of the considered stationary state, and $\hat{v}_k$ the Fourier of the two-body potential.
Note that since $1/p$ is not integrable close to $p=0$, \eqref{eq:Lambda_hom} may not be well defined for $\lambda=0$, and in any case is not differentiable
along the imaginary axis. In particular, the function defined by \eqref{eq:Lambda_hom} for $\lambda\in \mathbb{R}$ is not differentiable for $\lambda=0$. 
This is the technical counterpart of the strong resonance phenomenon between a non oscillating perturbation and particles with vanishing velocities. We come back now to the non homogeneous case.

\subsubsection{Eigenfunctions}
The eigenfunctions with eigenvalue $\lambda$ of $L_\mu$ and $L_\mu^{\dagger}$ (recall that $\lambda$ is real) are respectively
expressed as
\begin{equation}
    \Psi_{\lambda}
    = \sum_{k\in K}\sum_{m\in\mathbb{Z}}
    \dfrac{im\nabla_{J}F_{\mu}}{\lambda+im\Omega}
    c_{k,m}e^{im\theta} b_{k}
\end{equation}
and
\begin{equation}
    \widetilde{\Psi}_{\lambda}
    = - \sum_{k\in K} \sum_{m\in\mathbb{Z}}
    \dfrac{c_{k,m}}{\lambda-im\Omega}
    e^{im\theta} \tilde{b}_{k},
\end{equation}
where
the real vectors $b=(b_{k})$ and $\tilde{b}=(\tilde{b}_{k})$
depending on $\lambda$ must satisfy the conditions
\begin{equation}
  \label{eq:condition-b} 
  \Lambda_{\mu}(\lambda) b(\lambda) = 0,
  \quad
  ^T\tilde{b}(\lambda) \Lambda_{\mu}(\lambda) = 0.
\end{equation}
where the upper $T$ represents the transposition.
The assumption
that $\lambda$ is a nondegenerate eigenvalue
implies that $\dim{\rm Ker}\Lambda_{\mu}(\lambda)=1$.
Otherwise, there would be two linearly independent eigenfunctions for $\lambda$.

\subsubsection{Analysis of the $a_2$ coefficient}
From the above eigenfunctions
the $a_{2}(\lambda)$ term is written as 
\begin{equation}
    \begin{split}
        a_{2}(\lambda)
        & = - 2\pi \sum_{j\in K} \sum_{k\in K} \sum_{l\in K} 
        \tilde{b}_{j} b_{k} b_{l} \\
        & \times \sum_{m\in\mathbb{Z}} \sum_{n\in\mathbb{Z}} 
        \int \dfrac{c_{j,m+n}^{\ast}\varphi_{m,n}}{\lambda+i(m+n)\Omega} dJ \\
    \end{split}
    \label{eq:a2generic}
\end{equation}
where
\begin{equation}
    \begin{split}
        \varphi_{m,n} & = \left[ 
          in \nabla_{J}
          \left( \dfrac{im\nabla_{J}F_{\mu}c_{k,m}}{\lambda+im\Omega}
          \right) c_{l,n} \right. \\
        & \hspace*{2em}
        \left. - \dfrac{im\nabla_{J}F_{\mu}c_{k,m}}{\lambda+im\Omega}
          im\nabla_{J}c_{l,n}
        \right].
    \end{split}
\end{equation}
We study the $\lambda$ dependence of the $a_{2}(\lambda)$ term
in the limit $\lambda\to 0^{+}$.

First, we estimate the order of magnitude of $\tilde{b}$.
The amplitude of $b$ can be chosen arbitrarily
but $\tilde{b}$ cannot, since $\tilde{\Psi}_{\lambda}$
must satisfy the normalization condition, which is expressed as
\begin{equation}
    1 = - \sum_{k\in K} \sum_{l\in K} \tilde{b}_{l}
    (\Lambda_{\mu})'_{lk}(\lambda) b_{k} = -\langle \tilde{b}, \Lambda_\mu' (\lambda)b\rangle,
\end{equation}
with $\langle \cdot,\cdot\rangle$ the standard scalar product. Making use of the facts that the $(\Lambda_{\mu})_{kl}$ are regular around $\lambda=0^+$, 
the normalization condition reads
\begin{equation}
  \label{eq:normalization-inthelimit}
  \begin{split}
    - 1
    & = \langle \tilde{b}, \Lambda_\mu'(0) b_{0} \rangle \\
    & + \lambda \langle \tilde{b}, \Lambda_\mu''(0) b_{0}
      + \Lambda'_{\mu}(0)b_{1}  \rangle
    + O(\lambda^{2}),
  \end{split}
\end{equation}
where we assumed that $b$ can be expanded in $\lambda$:
\begin{equation}
  b(\lambda) = b_{0} + \lambda b_{1} + O(\lambda^{2}).
  \label{eq:b}
\end{equation}
We will show that the amplitude of $\tilde{b}(\lambda)$ diverges
in the limit $\lambda\to 0^{+}$, and accordingly,
we assume that $\tilde{b}(\lambda)$ can be expanded as
\begin{equation}
  \tilde{b}(\lambda) = \lambda^{-y} \tilde{Y}(\lambda),
  \quad
  \tilde{Y}(\lambda) = \tilde{b}_{0} + \lambda \tilde{b}_{1} + O(\lambda^{2}).
\end{equation}
The normalization condition \eqref{eq:normalization-inthelimit} together with \eqref{eq:b} imposes $y\geq 0$.
First, note that the definition of $b$ \eqref{eq:condition-b} reads, at order $\lambda$:
\begin{equation}
  \Lambda_{\mu}(0)b_{1} = - \Lambda_{\mu}'(0)b_{0}.
\end{equation}
Since $\Lambda_{\mu}(0)$ is symmetrical, this implies that $\Lambda_{\mu}'(0)b_{0}$ is
orthogonal to the kernel of $\Lambda_{\mu}(0)$. Now, writing at leading order 
the definition of $\tilde{b}$ \eqref{eq:condition-b}, we obtain $\Lambda_\mu(0)\tilde{b}_0=0$.
Hence, we conclude that $\langle \tilde{b}_{0}, \Lambda_\mu'(0)b_0\rangle=0$,
and the first term of \eqref{eq:normalization-inthelimit} is estimated as
\begin{equation}
  \langle \tilde{b}(\lambda), \Lambda_\mu'(0)b_0\rangle = O(\lambda^{1-y}).
\end{equation}
The second term of \eqref{eq:normalization-inthelimit} is
also estimated in the same ordering as
\begin{equation}
  \lambda \langle \tilde{b}, \Lambda_\mu''(0) b_{0}
    + \Lambda'_{\mu}(0)b_{1}  \rangle
  = O(\lambda^{1-y}),
\end{equation}
and the remaining terms are higher than $O(\lambda^{1-y})$.
Consequently, to satisfy the normalization condition
\eqref{eq:normalization-inthelimit}, 
the exponent $y$ should be unity and $\tilde{b}=O(1/\lambda)$.

In order to show that $\tilde{b}$ is the unique source
of divergence in the $a_{2}(\lambda)$ term,
we have to check two cases which may give an extra diverging factor $1/\lambda$
in the integrand of $a_{2}(\lambda)$ (see \eqref{eq:a2generic}): $m=0$ and $m+n=0$.
It is easy to find that the contribution from the former vanishes.
Expanding $a_{2}(\lambda)$,
we find that the contribution from the latter is canceled out
at leading order between $\pm m$,
and the integral gives a finite value.
Therefore, we conclude that $a_{2}(\lambda)$ is of order $1/\lambda$
in the limit $\lambda\to 0^{+}$.

\section{An explicit example: cosine potential (HMF model)}
\label{sec:cosine}
The previous computations, for a generic 1D potential, are a bit abstract, and complicated enough already at order $A^2$. 
In order to perform higher order computations, and to prepare for precise numerical tests, we specialize in this section
to a particular interaction potential $v(q)=-\cos q$, with $q\in [0,2\pi[$ and periodic boundary conditions; this model is sometimes called the Hamiltonian 
mean field (HMF) model
\cite{AntoniRuffo,ReviewCampaDauxoisRuffo}.
 All computations then become more explicit, which will be useful to analyze higher orders, see Sec.\ref{sec:higher_orders}. In addition, we can also make use in this context of the "rearrangement" approach described in \cite{OgawaYamaguchi15}: this will 
 provide an alternative analytical approach to test the results, see Sec.\ref{sec:rearrangement}.

\begin{center}
 \begin{figure}
      \includegraphics[width=0.48\textwidth]{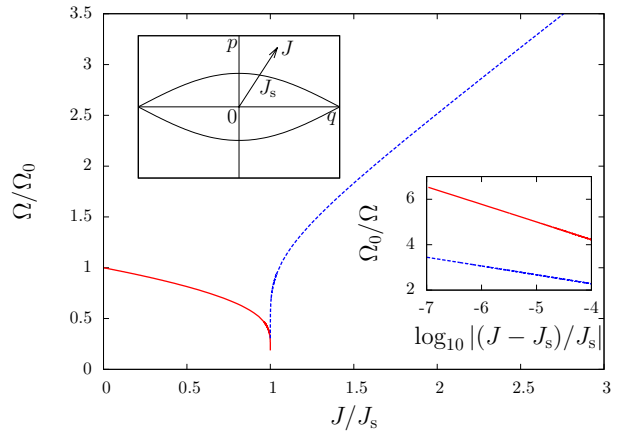}
   \caption{
     (Color online) Frequency $\Omega$ as a function of the action $J$ in the HMF model.
     The action is $J_{\rm s}$ at the separatrix
       (see the upper-left inset representing the phase space $(q,p)$),
       where $J_{\rm s}$ is different if the separatrix is approached from below or from above \cite{BOY10}.
     $\Omega$ goes to $\Omega_{0}$ in the limit $J\to 0^{+}$.
     The lower-right inset shows the logarithmic divergence of $1/\Omega$
     around the separatrix action $J_{\rm s}$
     for both inside (red solid upper line) and outside (blue broken lower line).
      }
    \label{fig:omega}
  \end{figure}
\end{center}
\subsection{Cosine potential: computation at order $A^2$}
The mean-field potential created by a phase space distribution $f$
is in this case $V_{MF}[f](q)=-M_{x}\cos q-M_{y}\sin q$ with
\[
M= M_x+i M_y = \iint f(q,p) e^{iq} ~dpdq.
\]
$M$ is usually called the "magnetization".
We choose $M_\mu$, the magnetization of $f_\mu$, to be real, and assume a symmetric initial condition, such that
 $M_y(t)=0$ for all $t$ (test simulations with $M_y(0)\neq0$ -not shown- showed a similar behavior).
We also assume $M_\mu\neq 0$, that is $f_\mu$ non homogeneous.
The one particle dynamics in $V_{MF}$ is a simple pendulum dynamics,
obviously integrable;
we introduce the associated action-angle variables $(J,\theta)$.

The biorthogonal functions in the expansion of the potential
are $\{\cos q,\sin q\}$ and the size of matrix $\Lambda_{\mu}(\lambda)$
is $2$ accordingly.
The matrix $\Lambda(\lambda)$ is diagonal: 
$\Lambda_{\mu}(\lambda)={\rm diag}(\Lambda_{\mu}^{c}(\lambda),\Lambda_{\mu}^{s}(\lambda))$ \cite{BOY10},
where
\begin{equation}
  \label{eq:spectrum-function}
  \Lambda_{\mu}^{c}(\lambda) = 1 + 2\pi \sum_{m\neq 0}
  \int \dfrac{imF'_{\mu}(J)|c_{m}(J)|^{2}}{\lambda+im\Omega(J)} dJ
\end{equation}
\begin{equation}
  \Lambda_{\mu}^{s}(\lambda) = 1 + 2\pi \sum_{m\neq 0}
  \int \dfrac{imF'_{\mu}(J)|s_{m}(J)|^{2}}{\lambda+im\Omega(J)} dJ
\end{equation}
and
\begin{equation}
  c_{m}(J) = \dfrac{1}{2\pi} \int_{0}^{2\pi} \cos q(\theta,J) e^{-im\theta} d\theta,
\end{equation}
\begin{equation}
  s_{m}(J) = \dfrac{1}{2\pi} \int_{0}^{2\pi} \sin q(\theta,J) e^{-im\theta} d\theta.
\end{equation}
The spectrum function satisfies $\Lambda_{\mu}^{s}(0)=0$
\cite{CampaChavanis2010},
which corresponds to the Goldstone mode.
The critical point $\mu_{\rm c}$ is, therefore, determined
by the other part and
the root of the equation $\Lambda_{\mu}^{c}(\lambda)=0$
tends to $0^{+}$ in the limit $\mu\to\mu_{\rm c}$.
Hereafter we consider the cosine part $\Lambda_{\mu}^{c}$ only,
and denote it by $\Lambda_{\mu}$ for simplicity.

In the above setting the eigenfunctions are 
\begin{eqnarray}
\label{eq:psi}
\Psi_\lambda &=& \sum_{m \in \mathbb{Z}} \frac{imF'_{\mu}(J)c_m(J)}{\lambda +im\Omega(J)}e^{im\theta} \\
\widetilde{\Psi}_\lambda &=& -\frac{1}{\Lambda_{\mu}'(\lambda)^{\ast}}\sum_{m\in \mathbb{Z}} \frac{c_m(J)}{\lambda-im\Omega(J)}e^{im\theta}.
\label{eq:psitilde}
\end{eqnarray}
It is easy to see that 
$\Lambda_{\mu}'(\lambda)\sim C\lambda$ for $\mu\to\mu_{\rm c}$,
inducing a divergence in $\widetilde{\Psi}_\lambda$. 
The possibility of resonances is readily seen on
Eqs.\eqref{eq:psi} and \eqref{eq:psitilde}:
they correspond to actions $J$ such that $\lambda \pm im\Omega(J)=0$.
Since we have assumed $\lambda \in \mathbb{R}$,
this can happen only at the separatrix $J=J_s$, where $\Omega(J_{\rm s})=0$
(see Fig.\ref{fig:omega}).
The divergence is very weak,
$\Omega(J)^{-1} = O(\ln|J-J_s|)$, thus
\emph{resonances are strongly suppressed}
since the number of particles around $J=J_{\rm s}$ is negligible.
On the other hand, for a complex $\lambda$ strong resonances are possible,
a priori similar to those occurring in the homogeneous case.

For the cosine potential (HMF model),
the above expression \eqref{eq:a2generic} for $a_{2}(\lambda)$ simplifies a little bit:
\begin{equation}
    \label{eq:a2-suppl}
    a_{2}(\lambda)
     = \dfrac{-2\pi }{\Lambda_{\mu}'(\lambda)}
     \int \sum_{m,n\in\mathbb{Z}}
     \dfrac{c_{m+n}^{\ast}\varphi_{m,n}(J)}{\lambda+i(m+n)\Omega} dJ
\end{equation}
where $\Lambda_{\mu}(\lambda)$ is given by \eqref{eq:spectrum-function}, and
\begin{equation}
    \begin{split}
        \varphi_{m,n}(J)
        =  m 
        \left[ n \dfrac{\partial}{\partial J} \left(
            \dfrac{F_{\mu}'c_{m}}{\lambda+im\Omega}
          \right) c_{n}
          - m \dfrac{F_{\mu}'c_{m}}{\lambda+im\Omega} c'_{n}
        \right]. \\
    \end{split}
\end{equation}
It is straightforward to see
that $a_{2}(\lambda)$ is of order $1/\lambda$ in the limit $\lambda\to 0^{+}$
due to the divergent factor $1/\Lambda_{\mu}'(\lambda)$.
Some details on how to compute numerically with a good accuracy
the function $\Lambda_{\mu}$ and its roots are given in the appendix.

\subsection{Computations at higher orders}
\label{sec:higher_orders}

We explain here how to compute the higher order terms. While the structure is standard, the effective computations are intricate. The first paragraph is valid for a general potential, but in order to explicitly work out the order of magnitude of $a_3$, we restrict to the cosine potential. We then show that $a_3\sim C/\lambda^3$, as $\lambda \to 0^+$.

\subsubsection{Structure of the computation for a general potential}
Let us parameterize the unstable manifold of $f_{\mu}$ as
\begin{equation}
  g = \sum_{k\geq 1} S_{k} A^{k},
  \quad
  S_{1} = \Psi_{\lambda}.
\end{equation}
Since the nonlinear part $N[g]$ of the Vlasov equation,
$\partial_{t}g=L_{\mu}g+N[g]$, is bilinear, we write it as
\[
  N[g] = B(g,g)~,~\mbox{with}~B(g,h)=\partial_Jg\partial_\theta V[g]-\partial_\theta g\partial_J V[g].
\]
Recalling that the reduced dynamics on the unstable manifold is
\begin{equation}
  \dot{A} = \sum_{k\geq 1} a_k A^k,
  \quad
  a_{1} = \lambda,
\label{eq:reduced}
\end{equation}
our goal is to provide a formal expression for $a_k$. This requires computing at the same time the $S_{k}$'s.
We write the time evolution equation for $g$ in two ways:
\begin{equation}
  \begin{split}
    \partial_{t} g
    & = \sum_{k\geq 1} \sum_{l\geq 1} A^{k+l-1} k a_{l} S_{k} \\
    & = \sum_{k\geq 1} A^{k} L_{\mu} S_{k}
    + \sum_{k\geq 1}\sum_{l\geq 1}A^{k+l}B(S_{k}, S_{l}), \\
  \end{split}
\end{equation}
where we have used \eqref{eq:reduced}.
Picking up the terms order by order, we have for any $k$
\begin{equation}
  (ka_{1}-L_{\mu})S_{k} = \sum_{l=1}^{k-1} \left[
    B(S_{k-l},S_{l}) - (k-l)a_{l+1}S_{k-l} \right].
\end{equation}
This equation for $k=1$ is simply $L_{\mu}\Psi_{\lambda}=\lambda\Psi_{\lambda}$,
and we focus on $k\geq 2$.

We now project these equations onto
${\rm span}\{\Psi_\lambda\}$ and ${\rm span}\{\Psi_\lambda\}^\perp$;
we note the corresponding projection operators 
$\Pi$ and $\Pi^\perp=\mathbb{I}-\Pi$,
where $\mathbb{I}$ is the identity.
The projection operators work as
\begin{equation}
  \begin{split}
    & (\Pi \circ L_{\mu}) \Psi_{\lambda} = \lambda \Psi_{\lambda},
    \quad
    (\Pi^{\perp}\circ L_{\mu}) \Psi_{\lambda} = 0, \\
    & (\Pi \circ L_{\mu}) S_{k} = 0,
    \quad
    (\Pi^{\perp}\circ L_{\mu})S_{k} = L_{\mu}S_{k},
    \quad k\geq 2.
  \end{split}
\end{equation}
The projection operator $\Pi$ induces
\begin{equation}
  \label{eq:ak}
  a_{k} \Psi_{\lambda} = \sum_{l=1}^{k-1} \Pi\circ B(S_{k-l},S_{l}),
\end{equation}
and $\Pi^{\perp}$ yields
\begin{equation}
  \label{eq:Sk}
  (k\lambda-L_{\mu})S_{k}
  = \sum_{l=1}^{k-1} \Pi^{\perp}\circ B(S_{k-l},S_{l})
  - \sum_{l=1}^{k-2} (k-l)a_{l+1}S_{k-l}.
\end{equation}
Equation \eqref{eq:ak} determines $a_{k}$ from $S_{l}$,
and \eqref{eq:Sk} $S_{k}$ from $S_{l}$ and $a_{l}$
with $l\in\{1,\cdots,k-1\}$.

\subsubsection{Cosine potential - Order of magnitude of $a_3$}
We now specialize to the cosine potential and focus on the third order coefficient $a_3$.

Calling $G_{k}$ the r.h.s. of Eq.\eqref{eq:Sk}, we have
\begin{equation}
  S_{k} = R(k\lambda) ~ G_{k},
\end{equation}
where $R(z) =(z-L_{\mu})^{-1}$ is the resolvent of $L_{\mu}$. Our first task is to determine the 
$\lambda$ dependence of $R(k\lambda) ~ G_{k}$. Let us then consider the equation\\
\[
(z-L_{\mu}) X = G~,
\]
and solve for $X(\theta,J)$.
Denoting the $m$-th Fourier component of $X$ in $\theta$ as $\hat{X}_{m}$,
we find after some computations, for all $m\in \mathbb{Z}$:
\begin{equation}
\hat{X}_m = \frac{\hat{G}_m}{z+im\Omega(J)} - C(z) \frac{imc_mF'_\mu(J)}{z+im\Omega(J)}
\label{eq:resolvent}
\end{equation}
where
\begin{equation}
  \label{eq:C-factor}
  C(z) = \dfrac{2\pi\sum_l \int \frac{\hat{G}_lc_l^\ast}{z+il\Omega(J')} dJ'}{\Lambda_\mu(z)}.
\end{equation}
Taking $z=2\lambda$ in \eqref{eq:resolvent}, we see that $\Lambda_\mu(2\lambda)^{-1}$ introduces a $1/\lambda^2$ divergence, and 
that the $l=0$ term in the sum yields an extra $1/\lambda$ divergence, unless $\hat{G}_{l=0}=0$. 
Thus, the $C$ factor \eqref{eq:C-factor} gives the leading singularity.

Now, recalling \eqref{eq:Sk}, we have to apply the resolvent to $G=\Pi^\perp B(\Psi_\lambda,\Psi_\lambda)$. 
Using the definition of $a_2$, we have
\begin{equation}
  S_2 = R(2\lambda)~B(\Psi_\lambda,\Psi_\lambda) - a_2 R(2\lambda)~\Psi_\lambda.
\label{eq:piperpB}
\end{equation}
We first note that $\widehat{B(\Psi_\lambda,\Psi_\lambda)}_{m=0}=0$
($B$ contains two terms, each containing a derivative with respect to $\theta$; hence the zeroth Fourier mode vanishes) and that $\hat{\Psi}_{\lambda,m=0}=0$. Hence the possible divergence related to $l=0$ in \eqref{eq:resolvent} does not exist, and the resolvent introduces only a $1/\lambda^2$ divergence.
We conclude that in the r.h.s. of \eqref{eq:piperpB}, the first term is $O(\lambda^{-2})$.
The second is a priori $O(\lambda^{-3})$, since $a_2\propto 1/\lambda$,
but we show now it is actually also $O(\lambda^{-2})$.
The $C$ factor \eqref{eq:C-factor} for $\widehat{(R(2\lambda)\Psi_{\lambda}})_{m}$ is
\begin{eqnarray}
C &=& \frac{2\pi\sum_m \int \frac{\hat{\Psi}_mc_m^\ast}{2\lambda+im\Omega(J)}dJ}{\Lambda_\mu(2\lambda)} \nonumber \\
&=& \frac{2\pi}{\Lambda_\mu(2\lambda)} \sum_m \int im \frac{F'_\mu(J) |c_m|^2}{(\lambda+im\Omega)(2\lambda+im\Omega(J))}dJ \nonumber \\
&=& \frac{2\pi}{\Lambda_\mu(2\lambda)} 6\lambda \sum_{m>0} \int m^2 \frac{F'_\mu(J) |c_m|^2}{|\lambda+im\Omega|^2 |2\lambda+im\Omega|^2}dJ\nonumber \\
&\sim & \frac{3}{2}\frac{ \Lambda_\mu'(\lambda)}{\Lambda_\mu(2\lambda)} = O(1/\lambda) \nonumber
\end{eqnarray}
where the last line is for $\lambda \to 0^+$.
Hence, the second term in the r.h.s. of \eqref{eq:piperpB} is $O(\lambda^{-2})$, and so is $S_2$. From \eqref{eq:ak}
\begin{equation}
  \label{eq:a3}
  a_{3} \Psi_{\lambda} = \Pi\circ \left[ B(S_{2},\Psi_\lambda) + B(\Psi_\lambda,S_2)\right],
\end{equation}
where the r.h.s. is $\Pi$ applied to a $O(\lambda^{-2})$.
Now, the projection $\Pi$ contains a diverging $1/\lambda$ factor, coming from the normalization factor $1/\Lambda_\mu'(\lambda)$, 
needed in $\widetilde{\Psi}_\lambda$
to ensure that $\langle \widetilde{\Psi}_\lambda,\Psi_\lambda \rangle =1$. Hence, except for a restricted set of functions $\varphi$ 
such that $\langle \widetilde{\Psi}_\lambda,\varphi \rangle=O(1)$, we have (for $\varphi$ independent of $\lambda$):
$\Pi \varphi \propto \frac{1}{\lambda}$. The exceptional $\varphi$ such that the projection $\Pi$ 
does not introduce a diverging $1/\lambda$ factor 
lie close to the kernels of $\Pi$ and $\Pi^\perp$ (the latter is just $\mbox{Span}(\Psi_\lambda)$).
With this in mind, it is not difficult to conclude that $a_3 \propto 1/\lambda^3$. This is the same divergence strength as the one which 
appears in the homogeneous case \cite{Crawford_prl}, although the mechanism inducing the divergence is different.

\subsection{Rearrangement formula for the cosine potential}
\label{sec:rearrangement}
The rearrangement formula is a powerful tool to predict
the asymptotic stationary state $\fA$
from a given initial state $\fI$. We show here that it allows to recover some of the 
above results, following a very different route.

The main result from the rearrangement formula is, roughly speaking,
\begin{equation}
    \fA = \ave{ \fI }_{\rm A},
\end{equation}
where the symbol $\ave{\cdot}_{\rm A}$ represents the average
over angle variable at fixed action, angle and action being associated with the asymptotic
Hamiltonian $\HA=H[\fA]$
whose potential part is determined self-consistently \cite{OgawaYamaguchi15}. 
We show that this formula predicts 
$|M_{\mu}^{\rm A}-M_{\mu}|=O(\lambda^{2})$,
where $M_{\mu}$ and $M_{\mu}^{\rm A}$ are magnetizations
in $f_{\mu}$ and $f_{\mu}^{\rm A}$ respectively:
this is consistent with the unstable manifold analysis.

We assume that $\fI=F_{\mu}(\HI)$.
The average $\ave{\cdot}_{\rm A}$ can be removed
when applied to any function of energy: $\ave{\varphi(\HA)}_{\rm A}=\varphi(\HA)$
by the definition \cite{OgawaYamaguchi15}.
Thus, we can expand $\ave{ \fI }_{\rm A}$
with respect to small $\delta H=\HA-\HI$, and have
\begin{equation}
    \fA = \fI + F_{\mu}'(\HI) \delta H 
    - F_{\mu}'(\HA) \ave{\delta H}_{\rm A} 
    + \cdots .
\end{equation}
Adding and subtracting $F'(\HI)\ave{\delta H}_{\rm I}$
where the average $\ave{\cdot}_{\rm I}$ is taken over
angle variable associated with $H_{\mu}$,
multiplying by $\cos q$ and integrating over $(q,p)$, we obtain
a self-consistent equation for $\delta M=M_{\mu}^{\rm A}-M_{\mu}$:
\begin{equation}
    \label{eq:rearrangement-self-consistent}
    \Lambda_{\mu}(0) \delta M + \mathcal{N}(\delta M) = 0,
\end{equation}
where $\mathcal{N}(\delta M)$ is of higher order
and we used the Parseval's equality \cite{Ogawa13}
to derive the coefficient $\Lambda_{\mu}(0)$.
Straightforward computations give
\begin{equation}
  \mathcal{N}(\delta M) = \left\{
    \begin{array}{ll}
      O((\delta M)^{3/2}) & (M_{\mu}=0) \\
      O((\delta M)^{2}) & (M_{\mu}\neq 0) \\
    \end{array}
  \right.
\end{equation}
where the homogeneous case is rather singular \cite{OgawaYamaguchi14},
and the leading nonzero solution to \eqref{eq:rearrangement-self-consistent} is
\begin{equation}
    \delta M = \left\{
      \begin{array}{ll}
          O(\Lambda_{\mu}(0)^{2}) & (M_{\mu}=0) \\
          O(\Lambda_{\mu}(0)) & (M_{\mu}\neq 0). \\
      \end{array}
    \right.
\end{equation}
The relation between $\Lambda_{\mu}(0)$ and the instability
rate $\lambda$ is obtained by
recalling that $\lambda$ satisfies $\Lambda_{\mu}(\lambda)=0$
and expanding $\Lambda_{\mu}(\lambda)$ with respect to $\lambda$.
This expansion implies
\begin{equation}
    \Lambda_{\mu}(0) = \left\{
      \begin{array}{ll}
          O(\lambda) & (\Lambda_{\mu}'(0)\neq 0)  \\
          O(\lambda^{2}) & (\Lambda_{\mu}'(0)=0 \text{ and } \Lambda_{\mu}''(0)\neq 0) \\
      \end{array}
    \right.
\end{equation}
and the former and the latter correspond to the homogeneous and
non homogeneous cases respectively.
Summarizing, the rearrangement formula gives the scaling 
$\delta M=O(\lambda^{2})$ for both homogeneous and non homogeneous cases,
although the mechanisms for these scalings are different in each case,
as already found in the unstable manifold reduction. 
We remark that, in the non homogeneous case, the leading three terms
of \eqref{eq:rearrangement-self-consistent} gives two nonzero solutions
of order $O(\lambda^{2})$ and $O(1)$,
whose stability is not clear by the rearrangement formula only.
The latter solution is a priori out of range of the method;
nevertheless, later
we will numerically observe that one direction of perturbation
drives the system to the $O(\lambda^{2})$ state,
and the other direction to $O(1)$.

\section{Direct numerical simulations}
\label{sec:simulations}
\begin{center}
  \begin{figure}
     \includegraphics[width=0.48\textwidth]{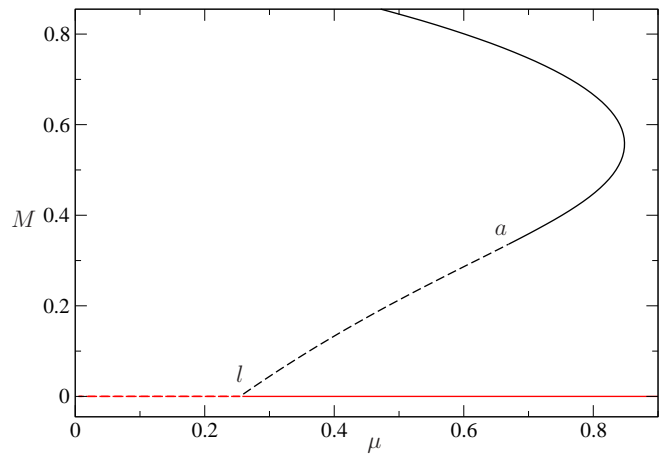}
    \caption{(Color online) Phase diagram of the Fermi distribution $F^0_\mu$ with $\beta=40$.
      Dotted lines correspond to unstable stationary solutions and solid lines to stable ones. We are interested in this article in the neighborhood of point $a$, where
      a branch of stable inhomogeneous $M_\mu \neq 0$ stationary solutions becomes unstable. At point $l$, a branch of stable homogeneous stationary solutions ($M_\mu=0$, red solid line) becomes unstable; we do not consider this bifurcation in this article.}
    \label{fig:M_mu}
  \end{figure}
\end{center}

\begin{center}
  \begin{figure}
     \includegraphics[width=0.48\textwidth]{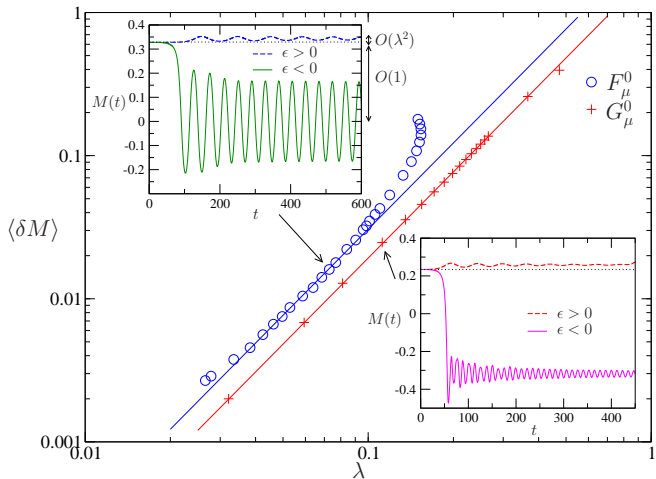}
    \caption{(Color online) $\left <\delta M\right >(\lambda)$ for $F_\mu^0$ (circles) and $G_\mu^0$ (crosses) with associated quadratic fit; $\beta=40$.
      For each function we show two runs of $M_x(t)=M(t)$ (here $M_y(t)=0$) with 
      positive and negative $\epsilon$.       $\left <\delta M\right >$ in the main diagram is computed as a long time average for $\epsilon>0$. For $F_\mu^0$, $|\epsilon|=1.8\cdot 10^{-6}$; for $G_\mu^0$, $|\epsilon|=10^{-5}$, except for $\lambda=0.032$ where $|\epsilon|=3\cdot 10^{-6}$. 
      Note that the reference stationary states $F_\mu^0$ follow the curve from point $a$ towards point $l$ on Fig.\ref{fig:M_mu}. Along this curve, $\lambda$ starts from $0$ at point $a$, then grows and reaches an upper limit, about $0.15$, before decreasing and reaching $0$ again at point $l$. This is why two values of $\mu$ may correspond to the same value of $\lambda$ (but different values of $\langle \delta M\rangle$), as can be seen around $\lambda=0.15$.  }
    \label{fig:deltaM}
  \end{figure}
\end{center}

Our analytical description of the bifurcation can be accurately tested
in the HMF case.
The time-evolved distribution function is obtained via a GPU parallel
implementation of a semi-Lagrangian scheme for the Vlasov HMF equation
with periodic boundary conditions~\cite{Rocha13}. 
We use a $2^n\times 2^n$ grid in position momentum
phase space truncated at $|p|= 2$
with $n$ up to $12$; the time step is usually $10^{-2}$. 
We use two families of reference stationary states
\begin{equation}
  F_{\mu}^{0}(\mathcal{E}) = N_{F}^{-1}/[1+e^{\beta (\mathcal{E}-\mu)}],
  \quad
  G_{\mu}^{0}(\mathcal{E}) = N_{G}^{-1}\mathcal{E}^2 e^{-\mu \mathcal{E}},
\end{equation}
with the one-particle energy 
$\mathcal{E}(q,p)=p^{2}/2+M_{\mu} (1-\cos(q))$,
where the magnetization $M_{\mu}$
has to be computed self-consistently,
and $N_{F}$ and $N_{G}$ are the normalization factors. Fig.\ref{fig:M_mu} presents $M_\mu$ and its bifurcations, for the family $F_{\mu}^{0}$.
For both families, a real positive eigenvalue appears
at a critical value of $\mu$: for $F_{\mu}^{0}$, this is at point $a$, see Fig.\ref{fig:M_mu}. 
The initial perturbation is
$\epsilon T(q,p)=\epsilon \cos(q)\exp(-\beta p^2/2)$.
Note that this is not proportional to the unstable eigenvector $\Psi_\lambda$:
this allows us to test the robustness of our unstable manifold analysis
with respect to the initial condition. For all simulation results presented in Fig.\ref{fig:deltaM}, the size of the perturbation $\epsilon$ was chosen small enough such that the saturated solution reached for $\epsilon>0$ does not depend on $\epsilon$. On the other hand, the smaller $\epsilon$, the more accurate computations are required to avoid numerical errors.
In particular, we have observed that numerical errors may drive the 
system far away from the reference stationary solution, following a dynamics similar to the one with $\epsilon <0$. In such cases, we have 
used a finer phase space grid: GPU computational power was crucial to reach very fine grids.
For example the initial Fermi distribution $F^0_\mu$ was very sensitive to numerical errors and to $\epsilon$: we took $\epsilon=1.8\cdot 10^{-6}$ with a $4096\times 4096$ grid; for $G^0_\mu$ distribution, which is much smoother, $\epsilon=10^{-5}$ and a $1024\times 1024$ grid was enough (except for the point corresponding to $\lambda=0.032$ where more precision was needed, and we took $\epsilon=3\cdot 10^{-6}$).

Typical evolutions for the order parameter $M(t)$ are shown in
Fig.\ref{fig:deltaM} (insets), for positive and negative $\epsilon$.
The asymmetry is clear: for one perturbation the change in
$\delta M=M(t)-M_{\mu}$ remains small, for the other it is $O(1)$.
In the case where $\delta M$ remains small, we compute its saturated
value by averaging the small oscillations;
the result is plotted as a function of $\lambda$ on Fig.\ref{fig:deltaM}:
the $\delta M\propto \lambda^2$ behavior is clear, for both families.
The fact that numerical simulations are able to reach this stationary state
suggests that it is a genuine stationary state, indeed stable with respect to the whole dynamics,
and not only on the unstable manifold.
However, longer simulations, or with smaller grid sizes (not shown),
indicate it is also easily destabilized by numerical noise.
We conclude that the $O(\lambda^{2})$ state is thus probably
close to the instability threshold. Notice that for the initial unstable reference state 
$|\Lambda_{\mu}(0)|=O(\lambda^2)$ is very small; hence the very close nearby stationary state with
$\delta M\propto \lambda^2$ may have a different stability,
although (in)stability of non homogeneous stationary states is rather robust
comparing with the homogeneous case
(see \cite{JMP} for a discussion).

We provide as Supplementary Material 3 videos, in order to better illustrate the phase space dynamics.
The video "Fermi\_eps\_+.mp4" shows the time evolution of the distribution function in phase space with initial condition $F^0_\mu(\mathcal{E}(q,p))+\epsilon T(q,p)$, for $\epsilon=+ 1.8\cdot 10^{-6}$, $\beta=40$, $(
M_{\mu}=0.328,~\mu=0.658)$ (it corresponds to the dashed blue curve in the upper inset of Fig.\ref{fig:deltaM}) from $t=0$ to $t=600$. Since the system reaches a new stationary state close to the original one, we observe almost no change in the distribution function. It is important to notice however that this picture is very different from that of the saturation of an instability over an homogeneous background: in that case, resonances would create small "cat's eyes" structures, which do not appear here.
To better appreciate the dynamics in this case, we also provide the video "Fermi\_eps\_+\_diff.mp4", which is the same as the previous one, except that the reference state $F^0_\mu$ has been substracted; hence the evolution of the perturbation is more clearly shown.

The video "Fermi\_eps\_-.mp4" shows the time evolution of the distribution function with the same parameter values except that $\epsilon=- 1.8\cdot 10^{-6}$ (it corresponds to the green curve in the upper inset of Fig.\ref{fig:deltaM}). This time the distribution changes completely its shape and seems to approach a periodic solution, far away from the original stationary distribution. In all simulations the relative error between the total energy of the system at a given time and the total initial energy is at most of the order of $10^{-7}$.

\section{Conclusion}
\label{sec:conclusion}
To summarize, we have elucidated for one-dimensional 
Vlasov equations the generic dynamics of a non oscillating instability
of a non homogeneous steady state. We have shown in particular:
i) due to the absence of resonances, the physical picture is
very different from the homogeneous case;
ii) a striking asymmetry: in one direction on the unstable manifold,
the instability quickly saturates, and the system reaches a nearby
stationary state, in the other direction the instability does not saturate,
and the dynamics leaves the perturbative regime;
iii) for an instability rate $\lambda$,
the nearby stationary state is at distance $O(\lambda^2)$
from the original stationary state.
Direct numerical simulations of the Vlasov equation with a cosine potential
have confirmed these findings,
and furthermore suggested that this phenomenology may have relevance
for initial conditions that are not on the unstable manifold.

These results are generic for 1D Vlasov equations. It will be of course important to assess
if they can be extended to two- and three-dimensional systems.
The structure of the reduced equation on the unstable manifold, which underlies the asymmetric behavior, should be valid in any dimension.
However, the crucial physical ingredient underlying our detailed computations leading to the $O(\lambda^2)$ scaling is the absence (or the weak effect) of 
resonance between the growing mode and the particles' dynamics. This has to be carefully 
analyzed in higher dimensions, and is left for a future study.
Let us note that the findings of \cite{Palmer90}
regarding the nonlinear evolution of the Radial Orbit Instability are in agreement with this picture: 
they start from a weakly unstable spherical state; close to this reference state, they find an axisymmetric weakly oblate stationary state; at finite distance 
from the reference state (outside the perturbative regime), they find a stable prolate stationary state. This is consistent with the asymmetric behavior we predict on the unstable manifold.
An open question regards the stability of the new weakly oblate stationary state: in our numerical examples,
it seems stable, but very close to threshold.
In \cite{Palmer90} they claim it is unstable, but since their numerical simulations
are much less precise, this could be a numerical artefact.
 
Finally, the possibility of an infinite dimensional reduced dynamics, as in \cite{DelCastilloNegrete}, 
remains open.
We have also restricted ourselves in this article
to a real unstable eigenvalue.
It is obviously worthwhile to investigate the bifurcation
with a complex pair of eigenvalues.
Resonances should appear in this case,
similar to those in the homogeneous case,
but with a different symmetry: new features are thus expected.

\acknowledgments
D.M. thanks T. Rocha for providing his GPU Vlasov-HMF solver.
Y.Y.Y. acknowledges the support of JSPS KAKENHI Grant Number 23560069.
J.B. acknowledges the hospitality of Imperial College, where part of this work was performed.

\section{Appendix: computing the function $\Lambda_{\mu}(\lambda)$ and its roots}
\label{sec:appendix}
Testing the predicted scaling $A_\infty \propto \lambda^2$, as done in Sec.\ref{sec:simulations}, requires computing $\lambda$ with a good accuracy. We present our method in this appendix. 
In order to compute the eigenvalue $\lambda$ associated with a given initial distribution $f_\mu(q,p)=F_\mu(J)$, one has to find a positive root of the dispersion function 
\begin{equation}
    \label{eq_supp:spectrum-function}
    \Lambda_{\mu}(\lambda) = 1 + 2\pi \sum_{m\neq 0}
    \int \dfrac{im F_{\mu}'(J)|c_{m}(J)|^{2}}{\lambda+im\Omega(J)} dJ.
\end{equation}
The two main numerical obstacles are to compute efficiently the functions $c_m(J)$, and the infinite sum in $m$. 
Fourier expansions of $\rm sn^2$ ($\rm sn$ is the sine Jacobi elliptic function) are known \cite{Milne2002}, and can be used with \cite{Ogawa13}
to obtain the following expressions:  
\begin{equation}
c_0(k)=
\begin{cases}
\dfrac{2 E(k)}{K(k)}-1, ~~~~~~~~~~~~~~~~~~\quad k<1
\\ \dfrac{2 k^2 E(1/k)}{K(1/k)}+1-2k^2, ~~~~\quad k>1
\end{cases}
\label{c0}
\end{equation}
and for $m\neq 0$ 
\begin{equation}
\begin{cases}
c_{2m}(k)=\dfrac{2\pi^2}{K(k)^2}\dfrac{mq(k)^m}{1-q(k)^{2m}},
\qquad ~~k<1
\\c_{2m+1}(k)=0,
\qquad ~~ \hfill k<1
\\c_{m}(k)=\dfrac{2\pi^2 k^2}{K(1/k)^2}\dfrac{mq(1/k)^m}{1-q(1/k)^{2m}},\quad k>1
\end{cases}
\label{cminf}
\end{equation}
where $K(k)$ and $E(k)$ are the elliptic functions of first and second kind respectively and $q(k)=\exp\left (-\pi K(\sqrt{1-k^2})/K(k)\right )$. Note that one may also use the $\rm sn \times dn$ and $\rm sn\times cn$ Fourier expansions to obtain similar expressions for the Fourier transform of $\sin q$ with respect to $\theta$.
These explicit expressions allow efficient computations of the $c_m$ and help in choosing a right truncation for the $m$ summation.
A standard numerical solver then provides precise values of $\lambda$.


\vspace*{1.5cm}
\begin{thebibliography}
\expandafter\ifx\csname natexlab\endcsname\relax\def\natexlab#1{#1}\fi
\expandafter\ifx\csname bibnamefont\endcsname\relax
  \def\bibnamefont#1{#1}\fi
\expandafter\ifx\csname bibfnamefont\endcsname\relax
  \def\bibfnamefont#1{#1}\fi
\expandafter\ifx\csname citenamefont\endcsname\relax
  \def\citenamefont#1{#1}\fi
\expandafter\ifx\csname url\endcsname\relax
  \def\url#1{\texttt{#1}}\fi
\expandafter\ifx\csname urlprefix\endcsname\relax\def\urlprefix{URL }\fi
\providecommand{\bibinfo}[2]{#2}
\providecommand{\eprint}[2][]{\url{#2}}

\bibitem{Elskens_book}
Y. Elskens and D. Escande,
Microscopic dynamics of plasmas and chaos IoP, 2003.

\bibitem{Bonifacio}
R. Bonifacio, C. Pellegrini and L. M. Narducci,
Collective instabilities and high-gain regime in a free electron laser,
\emph{Opt. Commun.} {\bf 50}, 373 (1984).

\bibitem{Picozzi_review}
A. Picozzi, J. Garnier, T. Hansson, P. Suret, S. Randoux, G. Millot, D.N. Christodoulides,
Optical wave turbulence: Towards a unified nonequilibrium thermodynamic formulation of statistical nonlinear optics,
\emph{Physics Reports} {\bf 542}, 1-132 (2014). 

\bibitem{Smereka}
P. Smereka,
A Vlasov equation for pressure wave propagation in bubbly fluids,
\emph{Journal of Fluid Mechanics}  {\bf 454}, 287-325 (2002).

\bibitem{Landau46}
L. Landau,
On the vibrations of the electronic plasma,
\emph{J. Phys.} (USSR) {\bf 10}, 25 (1946).

\bibitem{MouhotVillani}
C. Mouhot and C. Villani,
Landau damping,
\emph{J. Math. Phys.} {\bf 51}, 015204 (2010).

\bibitem{Balmforth_review}
N.J. Balmforth, P.J. Morrison, J.L. Thiffeault,
Pattern formation in Hamiltonian systems with continuous spectra; a normal-form single-wave model,
arXiv preprint arXiv:1303.0065, 2013.

\bibitem{Baldwin64}
D. E. Baldwin,
Perturbation Method for Waves in a Slowly Varying Plasma,
\emph{Phys. Fluids} {\bf 7}, 782 (1964).

\bibitem{ONeil65}
T. O'Neil,
Collisionless Damping of Nonlinear Plasma Oscillations,
\emph{Phys. Fluids} {\bf 8}, 2255 (1965).

\bibitem{Crawford}
J. D. Crawford,
Amplitude equations for electrostatic waves: Universal singular behavior in the limit of weak instability,
\emph{Phys. Plasmas} {\bf 2}, 97 (1995).

\bibitem{DelCastilloNegrete}
D. del-Castillo-Negrete,
Nonlinear evolution of perturbations in marginally stable plasmas,
\emph{Phys. Lett. A} {\bf 241}, 99 (1998).

\bibitem{Palmer90}
P. L. Palmer, J. Papaloizou and A. J. Allen,
Neighbouring equilibria to radially anisotropic spheres: possible end-states for violently relaxed stellar systems,
\emph{Mon. Not. R. astr. Soc.} {\bf 246}, 415 (1990).

\bibitem{BGK}
I. B. Bernstein, J. M. Greene and M. D. Kruskal,
Exact Nonlinear Plasma Oscillations,
\emph{Phys. Rev.} {\bf 108}, 546 (1957).

\bibitem{Schwarzmeier79} 
J. L. Schwarzmeier, H. R. Lewis, B. Abraham Shrauner and K. R. Symon,
Stability of Bernstein-Greene-Kruskal equilibria,
\emph{Phys. Fluids} {\bf 22}, 1747 (1979).

\bibitem{Manfredi00}
G. Manfredi and P. Bertrand,
Stability of Bernstein-Greene-Kruskal modes,
\emph{Phys. Plasmas} {\bf 7}, 2425 (2000).

\bibitem{Lin05}
Z. Lin,
Nonlinear instability of periodic BGK waves for Vlasov-Poisson system,
\emph{Commun. Pure Appl. Math.} {\bf 58}, 505 (2005).

\bibitem{Pankavitch14}
S. Pankavich and R. Allen,
Instability conditions for some periodic BGK waves in the Vlasov-Poisson system,
\emph{Eur. Phys. J. D} {\bf 68}, 363 (2014).

\bibitem{OgawaYamaguchi15}
S. Ogawa and Y. Y. Yamaguchi,
Landau-like theory for universality of critical exponents in quasistationary
states of isolated mean-field systems,
\emph{Phys. Rev. E} {\bf 91}, 062108 (2015).

\bibitem{Kalnajs77}
  A.J. Kalnajs,
  Dynamics of flat galaxies. IV - The integral equation for normal modes in matrix form,
  \emph{Astrophys. Journal} {\bf 212}, 637 (1977).

\bibitem{Lewis79}
  H.R. Lewis and K.R. Symon,
  Linearized analysis of inhomogeneous plasma equilibria: General theory,
  \emph{J. Math. Phys.} {\bf 20}, 413 (1979).

\bibitem{BOY11}
  J. Barr\'e, A. Olivetti and Y.Y. Yamaguchi,
  Algebraic damping in the one-dimensional Vlasov equation,
  \emph{J.~Phys.~A} {\bf 44}, 405502 (2011).

\bibitem{AntoniRuffo}
M. Antoni and S. Ruffo,
Clustering and relaxation in Hamiltonian long-range dynamics,
\emph{Phys. Rev. E} {\bf 52}, 2361 (1995).

\bibitem{ReviewCampaDauxoisRuffo}
A. Campa, T. Dauxois and S. Ruffo,
Statistical mechanics and dynamics of solvable models with long-range interactions,
\emph{Phys. Rep.} {\bf 480}, 57 (2009).

\bibitem{BOY10}
J. Barr\'e, A. Olivetti and Y. Y. Yamaguchi,
Dynamics of perturbations around inhomogeneous backgrounds in the HMF model,
\emph{J. Stat. Mech.} (2010) P08002.

\bibitem{CampaChavanis2010}
  A.Campa and P.H.Chavanis,
  A dynamical stability criterion for inhomogeneous quasi-stationary states in long-range systems,
  \emph{J. Stat. Mech.} (2010) P06001.

\bibitem{Crawford_prl}
  J.D. Crawford,
  Universal Trapping Scaling on the Unstable Manifold for a Collisionless Electrostatic Mode,
  \emph{Phys. Rev. Lett.} {\bf 73}, 656 (1994).

\bibitem{Ogawa13}
S. Ogawa,
Spectral and formal stability criteria of spatially inhomogeneous stationary solutions to the Vlasov equation for the Hamiltonian mean-field model,
\emph{Phys. Rev. E} {\bf 87}, 062107 (2013).

\bibitem{OgawaYamaguchi14}
S. Ogawa and Y. Y. Yamaguchi,
Nonlinear response for external field and perturbation in the Vlasov system,
\emph{Phys. Rev. E} {\bf 89}, 052114 (2014).

\bibitem{Rocha13}
T. M. Rocha Filho,
Solving the Vlasov equation for one-dimensional models with long range interactions on a GPU,
\emph{Comput. Phys. Comm.} {\bf 184}, 34 (2013).

\bibitem{JMP} J. Barr\'e and Y.Y. Yamaguchi, On the neighborhood of an inhomogeneous stable stationary solution of the Vlasov equation - Case of an attractive cosine potential, \emph{J.~Math. Phys.} {\bf 56}, 081502 (2015).

\bibitem{Milne2002}
  S.~C. {Milne},
  Infinite families of exact sums of squares formulas, Jacobi elliptic functions, continued fractions, and Schur functions,
  \emph{Springer}, (2002).


\end{thebibliography}
\end{document}